\documentstyle[aps,prb,epsfig]{revtex}
\begin{document}
\twocolumn
\wideabs{
\title{Electronic properties of quantum bars}
\author{I. Kuzmenko, S. Gredeskul, K. Kikoin,  Y. Avishai}
\address{Department of Physics, Ben-Gurion University of the Negev,
Beer-Sheva}
\date{\today}
\maketitle
\begin{abstract}
Spectrum of boson fields and two-point correlators 
are analyzed in a quantum bar system  
(a superlattice formed by two crossed interacting arrays of quantum 
wires), with short range 
interwire interaction.  The standard bosonization procedure is 
shown to be valid, within the two wave approximation.  The system behaves as 
a sliding Luttinger liquid in the vicinity of the $\Gamma $ point, but its 
spectral and correlation characteristics have either 1D or 2D 
nature depending on the direction of the wave vector in the rest of 
the Brillouin zone.  Due to interwire interaction, 
unperturbed states, propagating 
along the two arrays of wires, are always mixed, and the transverse components of 
the correlation functions do not vanish.  
This mixing is especially strong around the diagonals of the Brillouin zone, 
where 
transverse correlators have the same order of magnitude as 
the longitudinal ones.
\end{abstract}
} \vspace{\baselineskip} 
\section{Introduction and overview}
Diverse $D-1$ dimensional objects embedded in $D$
dimensional structures  were recently investigated experimentally and analyzed
theoretically.  Rubbers and various percolation networks are examples
of such disordered $D-1$ objects, whereas self-organized stripes in
oxicuprates, manganites, nanotube ropes and Quantum Hall systems are
examples of periodic structures of this kind.  In some cases, the
effective dimensions of such ordered structures may be intermediate
{\em e.g.} between $D=1$ and $D=2$.  They are especially promising 
candidates for
studying novel electronic correlation properties, which, 
in particular, are relevant for
the search of Luttinger liquid (LL) finger-prints in two dimensions.  This
challenging idea is motivated by noticing some 
unusual properties of electrons in Cu-O
planes in High-T$_c$ materials
\cite{Anders}.  However, the Fermi liquid state seems to be rather robust in
two dimensions.  In this respect, a 2D system of weakly coupled 1D
quantum wires \cite{Avr,Avi,Guinea} looks promising. Indeed,
a theoretical analysis of 
stable LL phases was recently presented for a system 
consisting of coupled parallel quantum
wires \cite{Luba00,Vica01,Silva01} and for $3D$ stacks of sheets of
such wires in parallel and crossed orientations \cite{Luba01}.  
In most
of these cases, the interaction between the parallel quantum wires is 
assumed to be perfect along the wire \cite{Luba01}, whereas the
interaction between the modes generated in different wires depends 
only on the inter-wire distance.  Along these lines, 
generalization of the LL theory for quasi $2D$ (and even
$3D$) systems is reported in Ref.  \onlinecite {Luba01} where the
interaction between two crossed arrays of parallel quantum wires
forming some kind of a network, depends on the distance between points
belonging to different arrays.  As a result, the grid of
crossed arrays retains its LL properties for the propagation along
both subsets of parallel wires, whereas cross-correlations remained
non-singular.  This LL structure 
can be interpreted as a quantum analog of a classical sliding
phases of coupled $XY$ chains\cite{Hern}.  A special case of 2D grid
where the crossed wires are coupled by tunneling interaction is
considered in Refs.  \onlinecite{Guinea,Castro} \\

\noindent In the present paper, a different course is elaborated.  We ask the
question whether it is possible to encode {\em both 1D and 2D
electron liquid regimes in the same system within the same energy
scale}.  In order to unravel the pertinent physics
 we consider a grid with
{\it short-range inter-wire interaction}.  This approximation might 
look shaky if applied for crossed stripe arrays in the cuprates. 
On the other hand, it seems
natural for $2D$ grids of nanotubes \cite{Egger,Rueckes}, or
artificially fabricated bars of quantum wires with grid periods
$a_{1,2}$ which exceed the lattice spacing of a single wire
or the diameter of a nanotube.  It will be demonstrated 
below that the short-range
interaction with radius $r_{0}\ll a_{1,2}$ turns out to be effectively
weak.  Therefore, such a quantum bar (QB) retains the $1D$ LL character
for the motion along the wires similarly to the case considered in 
Ref.\onlinecite{Luba01}.  At the same time, however, 
the boson mode propagation along the
diagonals of the grid is also feasible. This process is 
essentially a two-dimensional one,  as well as the shape of the Brillouin
zone and equipotential surfaces in the reciprocal QB lattice.\\

\noindent
Before developing the formalism, a few words about the 
main assumptions are in order. Our attention
here is mainly focused
on charge modes, so it is assumed that there is a gap for spin
excitations. Next, we are mainly interested in electronic
properties of QB which are not related to simple charge instabilities like
commensurate CDW, so that the (for simplicity equal) 
periods $a_{1}=a_{2}=a$ are
supposed to be incommensurate with the lattice spacing.  The Brillouin
zone (BZ) of the QB superlattice is two-dimensional, and the nature
of excitations propagating in this BZ is determined by Bragg
interference of modes with the superlattice wave vector.  This
interference (Umklapp processes) is, of course destructive for LL
excitations with both wave vector components close to multiple integers
of $2\pi/a$.  However, in case of weak scattering
$Vag/\hbar v_F\ll 1,$ where $V$ is the interwire interaction strength, 
$g$ is the usual 
LL parameter and $v_F$ is the Fermi velocity,
only two-wave interference processes near the boundaries of the BZ
are significant. One can then hope that the harmonic boson
modes survive in the major part of the BZ, and that the Hamiltonian of the
QB  might still be diagonalized without losing the main characteristic
features of the LL physics.

\section{Formalism: Hamiltonian and main approximations}
We consider a $2D$ periodic grid consisting of two arrays of $1D$
quantum wires of length $L$ oriented along unit vectors ${\bf e}_{1,2}$ with
an angle $\varphi$ between them.  The full Hamiltonian of the system is,
\begin{equation}
  H =  {H}_{1} + {H}_{2} +  H_{{int}}.
  \label{TotHam1}
\end{equation}
The Hamiltonian ${H}_{i}$ describes the 1D boson field
in the $i$-th array of wires ($i=1,2$)
\begin{eqnarray*}
{H}_{1} & = &
         \frac{\hbar{v}}{2}\sum_{{n}_{2}}
              \int\limits_{-L/2}^{L/2} {dx}_{1}
              \left\{
                   {g}{\pi}_{1}^{2} \left( x_1,
                   {n}_{2}a\right)+\right.\nonumber\\
               & &    \left.\frac{1}{g}
                   \left(
                        {\partial}_{{x}_{1}}
                        {\theta}_{1}
                        \left(
                             {x}_{1},{n}_{2}a
                        \right)
                   \right)^2
         \right\},
         \label{Eq.2}
\end{eqnarray*}
\begin{eqnarray*}
{H}_{2}& = &
     \frac{\hbar{v}}{2}\sum_{{n}_{1}}
         \int\limits_{-L/2}^{L/2}{dx}_{2}
         \left\{
          g{\pi}_{2}^{2}\left({n}_{1}a,{x}_{2}\right)+
              \right.\nonumber\\
             & &\left.\frac{1}{g}
              \left(
                   {\partial}_{x_2}
                   {\theta}_{2}
                   \left(
                        {n}_{1}a,{x}_{2}
                   \right)
              \right)^2
         \right\}.
         \label{Eq.3}
\end{eqnarray*}
Here $n_i$ enumerates the wires in the $i$-th array, $x_{i}$ is a 1D
contiuous coordinate along  ${\bf e}_{i}$, while $(\theta_i,\pi_i)$
are the conventional canonically conjugate boson fields (see, e.g.,
Ref.[\onlinecite {Delft}]).  The Fermi velocities $v_{F_{1,2}}=v$ and
the LL parameters $g_{1,2}=g$ are taken to be the same for both
arrays.  The modulus of each component of a quasimomentum in the
first BZ does not exceed $Q/2$ where $Q=2\pi/a$.  Generalization
to the case of different parameters $v_{i},g_{i} ,a_{i}$ is
straightforward.\\

\noindent
The interwire interaction results from a
 short--range contact capacitive 
coupling in
the crosses of the bar,
\begin{eqnarray}
H_{{int}} & = & \sum\limits_{{n}_{1},{n}_{2}}
            \int\limits_{-L/2}^{L/2} dx_1 dx_2
            V(x_1-n_1a,n_2a-x_2)\times \nonumber\\
           & & {\rho}_{1}({x}_{1},{n}_{2}a)
            {\rho}_{2}({n}_{1}a,{x}_{2}).
            \label{Eq.4}
\end{eqnarray}
Here $\rho_i({\bf r})$ are density operators, and $V({\bf r}_{1}-{\bf
r}_{2})$ is a short-range interwire interaction. Physically, 
it represents a highly screened Coulomb
electron--electron interaction  between two
charges located at the points ${\bf r}_{1}=(x_{1},n_{2}a)$ and ${\bf
r}_{2}=(n_{1}a,x_{2}).$ In what follows we use an interaction
of the form,
\begin{eqnarray*}
    V({\bf r})=\frac{V_0}{2}
        \Phi\left
        (\frac{x_1}{r_0},\frac{x_2}{r_0}
        \right).
    \label{Eq.5}
\end{eqnarray*}
Here $V_0$ is the coupling strength and $r_{0}$ is 
the screening radius.  The function
$\Phi(\xi_{1},\xi_{2})$ vanishes for $|\xi_{1,2}|\ge 1,$
and is normalized by condition ${\Phi}(0,0)=1.$ 
For simplicity it is taken to be an even function of its two
variables.  In terms of boson field operators
${\theta}_{i}$, the interaction acquires the following form,
\begin{eqnarray}
    H_{{int}} & = & V_{0}\sum\limits_{{n}_{1},{n}_{2}}
            \int\limits_{-L/2}^{L/2} dx_1 dx_2
            \Phi\left
        (\frac{x_1-n_1a}{r_0},\frac{n_2a-x_2}{r_0}
        \right)\times {}\nonumber\\
        && {} \partial_{x_1}\theta_1(x_1,n_2a)
         \partial_{x_2}\theta_2(n_2a,x_2).
    \label{Eq.6}
\end{eqnarray}

\noindent Anticipating a Fourier expansion we introduce a
a 2D basis of periodic Bloch functions which is constructed in terms
of 1D Bloch functions for non-interacting QB,
\begin{equation}
{\Psi}_{p,p',{\bf q}}({\bf r})={\psi}_{p,q_{1}}(x_{1})
            {\psi}_{p',q_{2}}(x_{2}),
            \label{Eq.7}
\end{equation}
where
\begin{eqnarray*}
{\psi}_{p,q}(x)=\frac{1}{\sqrt{L}}
            e^{iqx}f_p(q,x),
           \label{Eq.8}
\end{eqnarray*}
and
\begin{eqnarray*}
f_p(q,x)=\exp\left\{
                  i{\mbox{ sign }}(q)
                  \left(-1\right)^{p+1}
                  \left[\frac{p}{2}\right]
                  Qx
             \right\}.
             \label{Eq.9}
\end{eqnarray*}
Here $p,p'=1,2,\ldots,$ are the band numbers in a reduced BZ, and
${\bf q}=\{q_{1},q_{2}\}$ is the crystal 
quasimomentum, $(|q_{i}|\leq Q/2).$ In momentum representation
the full Hamiltonian (\ref{TotHam1}) then acquires the form,
\begin{eqnarray}
H & = & \frac{{\hbar}{v}{g}}{2a}\displaystyle{
                      \sum_{i=1}^{2}
                      \sum_{p}
                      \sum_{\bf q}}
                      {\pi}_{ip{\bf q}}^{+}
                      {\pi}_{ip{\bf q}}
                      +\nonumber\\
          & &\frac{\hbar}{2vga}\displaystyle{
          \sum_{ii'=1}^{2}
                 \sum_{pp'}
                 \sum_{\bf{q}}
                 }
                 {W}_{ipi'p'{\bf q}}
                 {\theta}_{i p {\bf q}}^{+}
                 {\theta}_{i'p'{\bf q}},
                 \label{TotHam2}
\end{eqnarray}
with matrix elements for interwire coupling given by,
\begin{eqnarray*}
{W}_{ipi'p'{\bf{q}}} =
    {\omega}_{i p {\bf{q}}}
    {\omega}_{i'p'{\bf{q}}}
    \left[
         {\delta}_{ii'}{\delta}_{pp'}+
         {\alpha}_{ipi'p'{\bf{q}}}
         \left(
              1-{\delta}_{ii'}
         \right)
    \right].
    \label{Eq.11}
\end{eqnarray*}
Here
\begin{equation}
{\omega}_{ip{\bf{q}}}=v
  \left(
       \left[\frac{p}{2}\right]Q+
       \left(-1\right)^{p+1}
       \left\vert{q}_{i}\right\vert
  \right),
  \label{dispersion}
\end{equation}
are eigenfrequencies of the ``unperturbed'' 1D mode pertaining to an
 array $i$,  band $p$ and quasimomentum ${\bf q}.$
The coefficients
\begin{equation}
{\alpha}_{ipi'p'{\bf{q}}}=\mbox{sign}{(q_{1}q_{2})}(-1)^{p+p'}
\frac{gV_{0}r_{0}^{2}}{{\hbar}va}
                       {\Phi}_{ipi'p'{\bf q}},
                       \label{alpha}
\end{equation}
are proportional to the dimensionless Fourier component of the
interaction strengths
\begin{eqnarray}
{\Phi}_{1p2p'{\bf{q}}} & =
  \int d{\xi}_{1}d{\xi}_{2}{\Phi}({\xi}_{1},{\xi}_{2})
  e^{-ir_0(q_1\xi_1+q_2\xi_2)}\times \nonumber\\
  & \times
  f_p^{*}(q_1,r_0\xi_1)
  f_{p'}^{*}(q_2,r_0\xi_2)={\Phi}_{2p'1p{\bf{q}}}.
  \label{Phi}
\end{eqnarray}

\noindent The Hamiltonian (\ref{TotHam2}) describes a system of coupled 
harmonic 
oscillators, which 
can be {\em exactly} diagonalized with the help of 
a certain  canonical linear transformation 
(note that due to spatial periodicity, it is 
already diagonal with respect to the quasimomentum ${\bf q}$).  The 
diagonalization procedure is, nevertheless,
 rather cumbersome due to the mixing of states 
belonging to different bands and arrays.  However, it is seen from 
Eq.(\ref{alpha}) that in the case $r_{0}\ll a$ the dimensionless 
interaction $\alpha$ becomes effectively weak (numerical 
estimates 
are given below) and a perturbation approach is applicable.  In this 
limit, the systematics of unperturbed levels and states is grossely conserved, 
at least in the low energy region corresponding to the 
first few bands.  Indeed, as it follows from the unperturbed dispersion 
law (\ref{dispersion}), the inter--band mixing is significant only 
along the high symmetry directions in the first BZ. The effect of  
interband interactions can be accounted for perturbatively in the rest of 
the BZ. Moreover, the mixing between modes within the same energy band is 
strong for waves with quasimomenta close to the diagonal of the first 
BZ. Away from the diagonal, this mixing effect  can also be calculated 
perturbatively.\\

\noindent
In second order of perturbation theory the above mentioned canonical 
transformation results in the following renormalized field operators 
and the corresponding renormalized eigenfrequencies for the first array:
\begin{eqnarray}
{\tilde\theta}_{1p{\bf{q}}} & = &
\left(1-\frac{1}{2}\beta_{1p{\bf q}}\right)
{\theta}_{1p{\bf{q}}}+\nonumber\\
&&+\sum\limits_{p'}
   \frac{{\alpha}_{1p2p'{\bf{q}}}
         {\omega}_{1p{\bf{q}}}
         {\omega}_{2p'{\bf{q}}}}
        {{\omega}_{1p{\bf{q}}}^{2}-
               {\omega}_{2p'{\bf{q}}}^{2}}
   {\theta}_{2p'{\bf{q}}},
   \label{theta1off}
\end{eqnarray}
where
\begin{equation}
    \beta_{1p{\bf q}} = \sum\limits_{p'}
   \left(
   \frac{{\alpha}_{1p2p'{\bf{q}}}
         {\omega}_{1p{\bf{q}}}
         {\omega}_{2p'{\bf{q}}}}
        {{\omega}_{1p{\bf{q}}}^{2}-
               {\omega}_{2p'{\bf{q}}}^{2}}\right
           )^{2},
               \label{beta}
\end{equation}
and
\begin{eqnarray*}
{\tilde\omega}_{1p{\bf{q}}}^{2} =
      {\omega}_{1p{\bf{q}}}^{2}
      \left[
      1+
       \sum\limits_{p'}
   \frac{{\alpha}^{2}_{1p2p'{\bf{q}}}
         {\omega}^{2}_{2p'{\bf{q}}}}
        {{\omega}_{1p{\bf{q}}}^{2}-
               {\omega}^{2}_{2p'{\bf{q}}}}
       \right].
\end{eqnarray*}
Corresponding formulas for the second array are obtained by
replacing $1p\rightarrow 2p$, and $1p'\rightarrow 2p'$.\\

\noindent
For quasimomenta lying off the diagonal of the first BZ, all
terms in these equations are non-singular.  Therefore, the
applicability of perturbation theory is related to the convergence
of the series on the right hand side of Eq.(\ref{beta}).  Away from
the first BZ boundary ($|{\bf q}|\ll Q/2$) the following estimation is
valid,
\begin{eqnarray*}
    \frac{{\omega}^{2}_{ip{\bf{q}}}}
     {{\omega}^{2}_{ip{\bf{q}}}-
      {\omega}^{2}_{{i'}1{\bf{q}}}}
     \approx 1+O\left(
                     \frac{q_{i'}^{2}}
                          {\left([p/2]Q\right)^{2}}
                \right),\\
    \label{Eq.19}
     (i,i')=(1,2),(2,1); \ \ p>1.\nonumber
\end{eqnarray*}
Therefore, the correction can be approximately estimated as
${\omega}_{1p{\bf{q}}}^{2}S_{{\bf q}}$ with
\begin{equation}
   S_{{\bf q}} =
\sum\limits_{p}
    {\alpha}_{112p{\bf q}}^{2} = \left(
      \frac
      {V_{0}gr_{0}^{2}}
      {\hbar v a}
      \right)^{2}
      \sum\limits_{p}
    {\Phi}_{112p{\bf q}}^{2}.
    \label{Sq}
\end{equation}
For a short-range interaction, $r_0\ll a$, ${\Phi}_{ipi'p'{\bf{q}}}$ 
is a
smooth function of $p$ and $p'$.  Therefore, the sum over $p$ in
Eq.(\ref{Sq}) can be replaced by an integral over the extended BZ with 
wave vector ${\bf{k}}$ whose components are
$$k_i=q_i+\mbox{sign }(q_i) \left(-1\right)^{p_i+1}
\left[p_i/2\right]Q.$$ For $|{\bf q}| \ll Q$ one gets
$S_{{\bf q}}=S_{0}(1+{\it o}(1))$ where,
\begin{eqnarray*}
    S_{0} = \left(\frac{V_{0}gr_{0}^{2}}{\hbar v a}\right)^{2}
    \frac{a}{2\pi}
\int dk {\tilde\Phi}_{0k}^{2},
    \label{Eq.22}
\end{eqnarray*}
and
\begin{eqnarray*}
    {\tilde\Phi}_{k_{1}k_{2}}=\int d\xi_{1}d\xi_{2} 
{\Phi}(\xi_1,\xi_2)
                      e^{-ir_0(k_1\xi_1+k_2\xi_2)}.
    \label{Eq.21}
\end{eqnarray*}
Finally, one arrives at the estimate,
\begin{equation}
    S_{0} \approx \left(\frac{V_{0} g \Phi_{0}}{\hbar v}\right)^{2}
\frac{r_{0}^{3}}{a},
    \label{Eq.22}
\end{equation}
where
\begin{eqnarray*}
    {\Phi}_{0}^{2}=\int d\xi d\xi_1 d\xi_2
\Phi(\xi_1,\xi)\Phi(\xi_{2},\xi).
    \label{Eq.23}
\end{eqnarray*}

\noindent In the case of QB formed by nanotubes, $V_{0}$ is the
Coulomb interaction screened at a distance of the order of the
nanotube radius\cite{Sasaki} $R_{0},$ so that $V_{0}\approx
e^{2}/R_{0}$ and $r_{0}\approx R_{0}$. Therefore,
\begin{equation}
    S_{0}= \left(\frac{e^{2} g \Phi_{0}}{\hbar v}\right)^{2}
\frac{R_{0}}{a}.
    \label{Eq.24}
\end{equation}
For carbon nanotubes one has \cite{Egger} $v \approx 8 \cdot 10^7$
cm/sec and $g\approx 1/3$.
 Then, $\Phi_{0}$ can be calculated e.g. for a Gaussian
model,
$$\Phi(\xi_{1},\xi_{2})\propto
e^{-(\xi_{1}^{2}+\xi_{2}^{2}-2\xi_{1}\xi_{2}\cos \varphi)}.$$
Taking for example $\varphi=\pi/3,$ one finds that interaction is 
effectively
weak if $a\gg 4 R_{0}.$ Therefore even slightly rarefied QB already
satisfies the desired condition. 
Thus, the dimensionless coupling constant $\alpha$ (\ref{alpha}) is a small
parameter of the theory.  Further calculations are carried out within an 
accuracy $o(\alpha^{2})$. Yet,
in the equations presented below, only the
first non-vanishing representative terms are retained 
(just for the sake of brevity). 

\noindent Consider now modes with quasi--momenta near the
diagonal of the first Brillouin zone, but away from its boundary.
In this case, the frequencies of the modes belonging to the same band 
coincide,
${\omega}_{1p{\bf q}}=
{\omega}_{2p{\bf q}}\equiv {\omega}_{p{\bf q}}.$
Therefore the modes are strongly mixed:
\begin{eqnarray}
{\tilde\theta}_{1p{\bf{q}}} & \approx &  \frac{1}{\sqrt{2}}
\left(
       {\theta}_{1p{\bf{q}}}+
       {\theta}_{2p{\bf{q}}}
  \right),\nonumber\\
{\tilde\theta}_{2p{\bf{q}}} & \approx & \frac{1}{\sqrt{2}}
  \left(
      -{\theta}_{1p{\bf{q}}}
      +{\theta}_{2p{\bf{q}}}
  \right).
 \label{thetaon}
\end{eqnarray}
The corresponding eigenfrequencies are shifted from their bare
values already in the first order in $\alpha$
\begin{eqnarray}
{\tilde\omega}_{1p{\bf{q}}}^{2} & \approx &
      {\omega}_{p{\bf{q}}}^{2}
       \left(
            1+
            {\alpha}_{1p2p{\bf{q}}}
       \right),\nonumber\\
{\tilde\omega}_{2p{\bf{q}}}^{2} & \approx &
      {\omega}_{p{\bf{q}}}^{2}
       \left(
            1-
            {\alpha}_{1p2p{\bf{q}}}
       \right).
       \label{Eq.26}
\end{eqnarray}
\begin{figure}[htb]
    \centering
    \includegraphics[width=80mm,height=78mm,angle=0,]{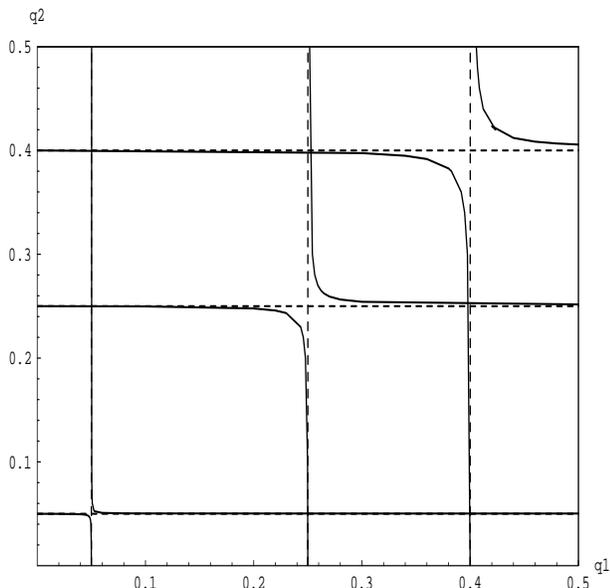}
    \epsfxsize=70mm 
\caption{Solid and dashed lines of 
equal energy for interacting and
noninteracting systems respectively.}
\label{equalenergy}
\end{figure}
\noindent These results show that the quantum states of the $2D$ quantum bar
conserve the quasi $1D$ character of the Luttinger--like liquid in the
major part of momentum space, and that the $2D$ effects can be
calculated within the framework of perturbation theory.  However, it
follows from Eqs.(\ref{thetaon}) that the bosons with quasimomenta
close to the diagonal of the first BZ are the strongly mixed bare 1D
bosons.  These excitations are essentially two-dimensional, and
therefore the lines of equal energy in this part of BZ are modified by
the 2D interaction (\ref{equalenergy}). It is
clearly seen that deviations from linearity occur only in a small
part of the BZ.
However, due to the absence of charge transfer, the Fermi surface of QB is still 
a ``cross'' obtained by superposition of two mutually perpendicular 
stripes of width $|2k_{F}|$ around the lines $k_{x}=0, k_{y} =0$ 
 (cf. Ref. \onlinecite{Guinea}).
The crossover from LL to FL behavior around isolated 
points of the BZ due to a single-particle hybridization (tunneling) 
for Fermi excitations was 
noticed in Refs. \onlinecite{Guinea,Castro}, where a mesh of 
horizontal and vertical stripes in superconducting cuprates was 
studied. 

\section{Correlations and observables}
The structure of the energy spectrum analyzed above predetermines
optical and transport properties of the QB. Let us consider an
optical conductivity ${\sigma}_{ii'}({\bf{q}},\omega)$ whose spectral
properties are given by a current--current correlator
\begin{equation}
{\sigma}_{ii'}({\bf{q}},
\omega)
={\rm Re}\left(
\frac{1}{\omega}
      \int\limits_{0}^{\infty}dt{e}^{i{\omega}t}
      \left\langle \left[
          {j}_{i1{\bf{q}}}(t),{j}_{i'1{\bf{q}}}^{\dag}(0)
      \right] \right\rangle\right).
      \label{CorrFunc}
\end{equation}
Here ${j}_{ip{\bf{q}}}=\sqrt{2}vg{\pi}_{ip{\bf{q}}}$ is a current
operator. For simplicity we restrict ourselves to the
first band. For non-interacting wires, the current-current 
correlator
is reduced to the conventional LL expression \cite{Voit},
\begin{equation}
\left\langle \left[
     {j}_{i1{\bf{q}}}(t),{j}_{i'1{\bf{q}}}^{\dag}(0)
\right] \right\rangle_0=
  -2ivg{\omega}_{i1{\bf{q}}}
  \sin({\omega}_{i1{\bf{q}}}t)
  {\delta}_{ii'}
 \label{29}
\end{equation}
with a metallic Drude peak
\begin{eqnarray*}
{\sigma}_{ii'}({\bf{q}},\omega>0)=
{\pi}vg
\delta({\omega}-{\omega}_{i1{\bf{q}}})
{\delta}_{ii'}.
\label{Drude_peak}
\end{eqnarray*}
For interacting wires,
where 
${\alpha}_{ipi'p'{\bf{q}}}\neq 0$, the
correlators may be easily calculated after diagonalization of the
Hamiltonian (\ref{TotHam2}) by the transformations (\ref{theta1off})
for ${\bf q}$ off the diagonal of the first BZ, or by the
transformation (\ref{thetaon}) for ${\bf q}$
lying on the diagonal of the BZ.\\

\noindent Consider first the optical conductivity for ${\bf{q}}$ far from the
diagonal of the first BZ. In this case, the transformations for the
field momenta can be obtained in a similar manner to the
transformations (\ref{theta1off}) for the field coordinates.  As a
result, one has:
\begin{eqnarray}
& \left\langle\left[
                 {j}_{11{ \bf{q}}}(t),
                 {j}_{11{\bf{q}}}^{\dag}(0)
  \right]\right\rangle  \approx
  & \nonumber \\
  & -2ivg\left(1-{\beta}^{2}_{\bf{q}}\right)
      {\tilde\omega}_{11{\bf{q}}}
      \sin({\tilde\omega}_{11{\bf{q}}}t)-&
\nonumber \\
&  -
 \displaystyle{
 2{i}{v}{g}
 \beta^{2}_{{\bf q}}
      {\tilde\omega}_{2p{\bf{q}}}
      \sin({\tilde\omega}_{2p{\bf{q}}}t),
 }&
\label{cur_cur}
\end{eqnarray}
\begin{eqnarray*}
  &
\left\langle\left[
                 {j}_{11{ \bf{q}}}(t),
                 {j}_{21{\bf{q}}}^{\dag}(0)
\right]\right\rangle  \approx
  -2i{v}{g}\beta_{{\bf q}}\times
  & \nonumber\\
  & \times \left(
     {\tilde\omega}_{11{\bf{q}}}
     \sin({\tilde\omega}_{11{\bf{q}}}t)-
     {\tilde\omega}_{21{\bf{q}}}
     \sin({\tilde\omega}_{21{\bf{q}}}t)
 \right), &
 \label{Eq.31-32}
\end{eqnarray*}
where $\alpha_{{\bf q}}\equiv \alpha_{1121{\bf q}}$ and
$$
\beta_{{\bf q}}=
\frac{\alpha_{{\bf q}} \omega_{11{\bf q}} \omega_{21{\bf q}}}
{\omega_{11{\bf q}}^{2}-\omega_{21{\bf q}}^{2}}.
$$
Then one obtains
\begin{eqnarray}
& {\sigma}_{11}({\bf{q}},\omega>0)  \approx
    {\pi}{v}{g}
    \left(
         1-{\beta}^{2}_{\bf{q}}
    \right)
    \delta
    \left(
         {\omega}-
         {\tilde\omega}_{11{\bf{q}}}
    \right) + & \nonumber\\
 &  +
    {\pi}{v}{g}
       \beta^{2}_{{\bf q}}
            \delta
            \left(
                 {\omega}-
                 {\tilde\omega}_{2p{\bf{q}}}
            \right),
 &
       \label{opt_ln}
\end{eqnarray}
\begin{eqnarray}
 &  {\sigma}_{12}({\bf{q}},\omega>0)  \approx
    {\pi}{v}{g}
    \beta_{{\bf q}}
    \times & \nonumber\\
  &  \times \Bigl[
         \delta
         \left(
              {\omega}-
              {\tilde\omega}_{11{\bf{q}}}
         \right)-
         \delta
         \left(
              {\omega}-
              {\tilde\omega}_{21{\bf{q}}}
         \right)
    \Bigr]. &
    \label{opt_tn}
\end{eqnarray}
The longitudinal optical conductivity (\ref{opt_ln}) (i.e. the
 conductivity within a given set of wires) has its main peak at the
frequency ${\tilde\omega}_{11{\bf{q}}}\approx{v\vert{q_1}\vert}$,
corresponding to the first band of the pertinent array, and an additional
weak peak at the frequency
${\tilde\omega}_{21{\bf{q}}}\approx{v\vert{q_2}\vert}$, corresponding
to the first band of a complementary array.  It 
contains also a set of weak peaks at frequencies
${\tilde\omega}_{2p{\bf{q}}}\approx [p/2]vQ$ ($p=2,3,\ldots$), omitted
in Eq.(\ref{opt_ln}) and corresponding to the contribution from 
higher bands of the complementary array.  At the same time, a 
second observable becomes relevant, namely, the
transverse optical  conductivity (\ref{opt_tn}).  It is
proportional to the interaction strength and has two peaks at
frequencies ${\tilde\omega}_{11{\bf{q}}}$ and 
${\tilde\omega}_{21{\bf{q}}}$ in the
first bands of both sets of wires.  For
$\left\vert{\bf{q}}\right\vert\to 0$,  Eq.(\ref{cur_cur}) reduces to
that for an array of noninteracting wires, and the transverse optical
 conductivity vanishes.\\

\noindent
In case where the quasimomenta ${\bf{q}}$ belong to the diagonal
of the first BZ, the transformations for the field momenta 
are similar in form to Eqs. (\ref{thetaon}).
The current-current correlation functions have the form
\begin{eqnarray}
&\left\langle\left[
                 {j}_{11{ \bf{q}}}(t),
                 {j}_{11{\bf{q}}}^{\dag}(0)
\right]\right\rangle   \approx   \nonumber\\
 & -i{v}{g}
   \left[
        {\tilde\omega}_{11{\bf{q}}}
        \sin({\tilde\omega}_{11{\bf{q}}}t)+
        {\tilde\omega}_{21{\bf{q}}}
        \sin({\tilde\omega}_{21{\bf{q}}}t)
   \right],
\label{Eq.35}
\end{eqnarray}
\begin{eqnarray*}
&\left\langle\left[
                 {j}_{1{ \bf{q}}}(t),
                 {j}_{2{\bf{q}}}^{\dag}(0)
\right]\right\rangle  \approx \nonumber\\
  &-ivg
\left(
     {\tilde\omega}_{11{\bf{q}}}
     \sin({\tilde\omega}_{11{\bf{q}}}t)-
     {\tilde\omega}_{21{\bf{q}}}
     \sin({\tilde\omega}_{21{\bf{q}}}t)
\right),
\label{}
\end{eqnarray*}
and the optical  conductivity is estimated as,
\begin{equation}
{\sigma}_{11}({\bf{q}},\omega>0)  \approx
    \displaystyle{\frac{{\pi}vg}{2}}
    \Bigl[
         \delta
         \left(
              {\omega}-
              {\tilde\omega}_{11{\bf{q}}}
         \right)+
         \delta
         \left(
              {\omega}-
              {\tilde\omega}_{21{\bf{q}}}
         \right)
    \Bigr],
       \label{opt_ld}
\end{equation}
\begin{eqnarray}
{\sigma}_{12}({\bf{q}},\omega>0) \approx
    \frac{{\pi}vg}{2}
    \Bigl[
         \delta
         \left(
              {\omega}-
              {\tilde\omega}_{11{\bf{q}}}
         \right)-
         \delta
         \left(
              {\omega}-
              {\tilde\omega}_{21{\bf{q}}}
         \right)
    \Bigr].
    \label{opt_td}
\end{eqnarray}
Here, the index $i$ of the array is omitted because the frequencies for
both arrays coincide, ${\omega}_{1p{\bf{q}}}={\omega}_{2p{\bf{q}}}.$
The longitudinal optical  conductivity (\ref{opt_ld}) has a split
double peak at frequencies ${\tilde\omega}_{11{\bf{q}}}$ and
${\tilde\omega}_{21{\bf{q}}},$ instead of a single peak.  Again, a
series of weak peaks (omitted in the r.h.s. of Eq.(\ref{opt_ld})) 
occurs
at frequencies ${\omega}_{p{\bf{q}}}$ corresponding to
contribution from higher bands $p=2,3,4,\ldots$.  The transverse
optical  conductivity (\ref{opt_td}), similarly to the nondiagonal case
(\ref{opt_tn}), has a split double peak at frequencies
${\tilde\omega}_{11{\bf{q}}}$ and ${\tilde\omega}_{21{\bf{q}}}$.\\

\noindent One of the main effects specific for a QB is the appearance of
non-zero transverse momentum--momentum correlation function. In
space-time coordinates $({\bf{x}},t)$ its representation reads,
\begin{equation}
G_{12}({\bf x},t) =
           \left\langle \left[
                             {\pi}_{1}(x_1,0;t),
                             {\pi}_{2}(0,x_2;0)
           \right] \right\rangle.
\end{equation}
This function describes the momentum response at the point $(0,x_{2})$
of the second array at the moment $t$ caused by initial
($t=0$) perturbation at the point $(x_{1}0)$ of the first array.
Standard calculations similar to those described above, lead to the following
expression,
\begin{eqnarray*}
& {G}_{12}({\bf{x}};t) =
                     \displaystyle{
                     \frac{V_0r_0^2}{2{\pi}{\hbar}{v}{a}}
                     \int\limits_{0}^{\infty} dk_1dk_2
                     }
                     {\tilde\Phi}_{k_1k_2}k_1k_2
                     \times &
                     \nonumber \\
                     & \times
                     \sin(k_1x_1)
                     \sin(k_2x_2)
                     \displaystyle{
                     \frac{k_2\sin(k_2vt)-
                           k_1\sin(k_1vt)}
                          {k_2^2-k_1^2}.
                     } &
              \label{spat_trans}
\end{eqnarray*}
\begin{figure}[htb]
	\centering
	\includegraphics[width=80mm,height=78mm,angle=0,]{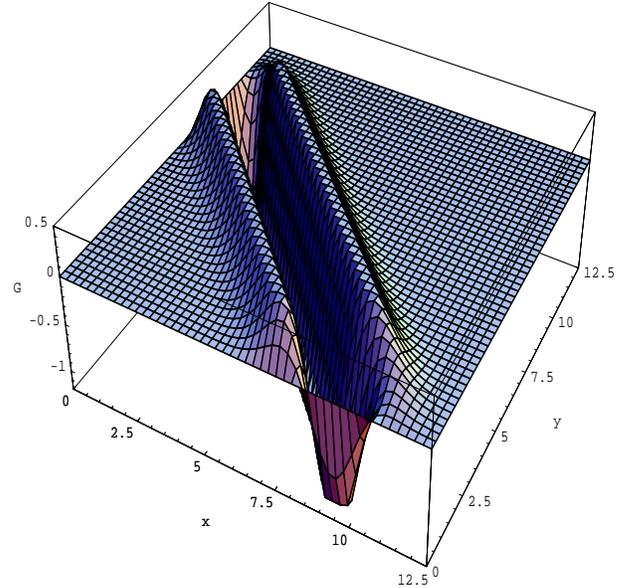}
	\epsfxsize=70mm \caption{The transverse correlation function
	$G_{12}(x_1,x_2;t)$ for $r_0=1$ and $vt=10$.}
\label{G12}
\end{figure}
\noindent This correlator is shown in Fig \ref{G12}
for $x_{1,2}>0$. Here the non-zero response corresponds to the line 
determined by the obvious kinematic condition $x_{1}+x_{2}=vt.$
The finiteness of the interaction radius slightly spreads this peak
and changes its profile.\\

\noindent Further manifestation of the 2D character of QB system is
related to a possible periodic energy transfer between the 
two arrays of wires.
Consider an initial perturbation which, in the system of {\it
non}-interacting arrays, corresponds to a plane wave propagating
within the first array
along the ${\bf e_{1}}$ direction,
\begin{eqnarray}
  \langle
  {\theta}_{1}(x_1,n_2a;t)
  \rangle & = &
  \frac{v{\rho}_{0}}{\sqrt{2}{\omega}_{11{\bf{q}}}}
  \sin(q_1x_1+q_2n_2a-{\omega}_{11{\bf{q}}}t),\nonumber\\
\langle
  {\theta}_{2}(n_{1}a,x_{2};t)
  \rangle & = & 0,
\end{eqnarray}
($\rho_{0}$ is the charge density amplitude).  If the wave vector
${\bf q},$ satisfying the condition $|{\bf{q}}|<<Q/2,$ is not close to
the first BZ diagonal, weak interwire interaction $\alpha$ slightly
changes the $\langle{\theta}_{1}\rangle$ component and leads to
the appearance of a small $\langle{\theta}_{2}\rangle\sim\alpha$
component.  But for ${\bf q}$ lying on the diagonal, both
components within the main approximation have the same order of magnitude
\begin{eqnarray}
   {\theta}_{1}(x_1,n_2a;t) & = &
   \frac{v\rho_0}{\sqrt{2}{\omega}_{1\bf{q}}}
   \cos\left(
                 \frac{1}{2}
                 {\alpha}_{\bf{q}}
                 {\omega}_{1{\bf{q}}}t
            \right)\nonumber\\
   & \times &\sin(q_1x_1+q_2n_2a-{\omega}_{1{\bf{q}}}t),
   \label{theta1d}
\end{eqnarray}
\begin{eqnarray*}
   {\theta}_{2}(n_1a,x_2;t) & = &
   \frac{v\rho_0}{\sqrt{2}{\omega}_{1\bf{q}}}
  \sin\left(
                 \frac{1}{2}
                 {\alpha}_{\bf{q}}
                 {\omega}_{1{\bf{q}}}t
            \right)\nonumber\\
        & \times &
        \cos(q_1n_1a+q_2x_2-{\omega}_{1{\bf{q}}}t).
\label{theta2d}
\end{eqnarray*}
This corresponds to a 2D propagation of a plane wave with wave
vector ${\bf q },$  {\it
modulated} by a ``slow'' frequency $\sim\alpha\omega.$ As a result,
an energy is periodically transfered from one array to another during a long
period $T\sim (\alpha\omega)^{-1}.$ These peculiar ``Rabi
oscillations'' may be considered as one of the fingerprints of 
the physics exposed in QB systems.

\section{Conclusion}
In conclusion, we have demonstrated that the energy spectrum of QB
shows the characteristic properties of LL at $|q|,\omega\to 0,$ but at
finite ${\bf q}$ the density and momentum waves may have either 1D or
2D character depending on the direction of the wave vector.  Due to
interwire interaction, unperturbed states, propagating along the two
arrays are always mixed, and transverse components of correlation
functions do not vanish.  For quasi-momentum lying on the
diagonal of the Brillouin zone, such a mixing is strong and transverse
correlators have the same order of magnitude as the longitudinal ones.\\
\section{Acknowledgement}
S.G. and K.K. are indebted to L.Gorelik, M. Jonson, I. Krive, and R.
Shekhter for numerous helpful discussions.  They also thank Chalmers
Technical University, where this work was started, for hospitality 
and support. This research is supported in part by grants feom the 
Israeli Science foundations the DIP German Israeli cooperation 
program and the U.S Israel BSF program.\\ 


\begin{thebibliography}{99}
\bibitem{Anders} P.W. Anderson, Science, {\bf235}, 1196 (1987).
\bibitem{Avr} J.E. Avron, A. Raveh, and B. Zur, Rev. Mod.Phys. {\bf 
60},
873 (1988).
\bibitem{Avi} Y. Avishai, and J.M. Luck, Phys. Rev. B{\bf 45}, 1074 
(1992).
\bibitem{Guinea} F. Guinea, and G. Zimanyi, Phys. Rev. B {\bf 47}, 
501 (1993). 
\bibitem{Luba00} V. Emery, E. Fradkin, S.A. Kivelson, and T.C. 
Lubensky,
Phys. Rev. Lett. {\bf 85}, 2160 (2000).
\bibitem{Vica01} A. Vishwanath and D. Carpentier,
Phys. Rev. Lett. {\bf86}, 676 (2001).
\bibitem{Silva01} J. Silva-Valencia, E. Miranda, and R.R. dos Santos,
cond-mat/0107114.
\bibitem{Luba01} R. Mukhopadhyay, C.L. Kane, and T.C. Lubensky,
Phys. Rev. B{\bf 63}, 081103(R) (2001); cond-mat/0102163.
\bibitem{Hern} C.S. Hern, T.C. Lubensky, and J. Toner, Phys. Rev. 
Lett. 
{\bf 83}, 2745 (1999). 
\bibitem{Castro} A.H. Castro Neto, and F. Guinea, Phys. Rev. Lett.
{\bf 80}, 4040 (1998).
\bibitem{Egger} R. Egger et. al., cond-mat/0008008.
\bibitem{Rueckes} T.Rueckes et. al., Science {\bf 289}, 94 (2000).
\bibitem{Delft} J. von Delft, and H. Schoeller,
Ann. der Physik {\bf 7}, 225 (1998).
\bibitem{Sasaki} K. Sasaki, cond-mat/0112178.
\bibitem{Voit} J. Voit, Rep. Prog. Phys., {\bf 58}, 977 (1994).
\end{thebibliography}
\end{document}